\documentclass[letterpaper,twocolumn,10pt]{article}

\makeatletter
\if@twocolumn\else\input twocolumn.sty\fi
\usepackage{mathptmx}  
\usepackage[T1]{fontenc}
\usepackage[utf8]{inputenc}
\usepackage{pslatex}

\usepackage[kerning,spacing]{microtype} \usepackage{flushend} 
\usepackage{cite}               \usepackage{breakurl}           \usepackage{url}                \usepackage{xcolor}             \usepackage[]{hyperref}         \hypersetup{                    colorlinks,
  linkcolor={green!80!black},
  citecolor={red!70!black},
  urlcolor={blue!70!black}
}

\def\maketitle{\par
 \begingroup
   \renewcommand\thefootnote{\fnsymbol{footnote}}\def\@makefnmark{\hbox to\z@{$\m@th^{\@thefnmark}$\hss}}\long\def\@makefntext##1{\parindent 1em\noindent
            \hbox to1.8em{\hss$\m@th^{\@thefnmark}$}##1}\if@twocolumn
     \twocolumn[\@maketitle]\else \newpage
     \global\@topnum\z@
     \@maketitle \fi\@thanks
 \endgroup
 \setcounter{footnote}{0}\let\maketitle\relax
 \let\@maketitle\relax
 \gdef\@thanks{}\gdef\@author{}\gdef\@title{}\let\thanks\relax}

\def\@maketitle{\newpage
 \vbox to 2.5in{
 \vspace*{\fill}
 \vskip 2em
 \begin{center}{\Large\bf \@title \par}\vskip 0.375in minus 0.300in
  {\large\it
   \lineskip .5em
   \begin{tabular}[t]{c}\@author
   \end{tabular}\par}\end{center}\par
 \vspace*{\fill}
}
}

\def\abstract{\begin{center}{\large\bf \abstractname\vspace{-.5em}\vspace{\z@}}\end{center}}

\def\section{\@startsection {section}{1}{\z@}{-3.5ex plus-1ex minus
    -.2ex}{2.3ex plus.2ex}{\reset@font\large\bf}}

\usepackage{tikz}
\usepackage{amsmath}

\usepackage{filecontents}

 \usepackage{array}
\usepackage{balance}
\usepackage{boldline}
\usepackage{calc}
\usepackage{cite}
\usepackage{dcolumn}
\usepackage{diagbox}
\usepackage{graphicx}
\usepackage{hhline}
\usepackage{listings}
\usepackage{makecell}
\usepackage{multirow}
\usepackage{url}
\usepackage{xcolor}
\usepackage{xspace}
\usepackage{adjustbox}

\hypersetup{
   urlcolor={blue!70!black}
}

\newcommand{\name}{\textit{HyCoR}\xspace}
\newcommand{\namese}{\textit{HyCoR-SE}\xspace}
\newcommand{\namele}{\textit{HyCoR-LE}\xspace}

\newcolumntype{d}[1]{D{.}{.}{#1}}

  \makeatother
\begin{document}

\title{HyCoR: Fault-Tolerant Replicated Containers Based on Checkpoint and Replay}

\author{
{\rm Diyu\ Zhou}\\
Computer Science Department, UCLA\\
zhoudiyu@cs.ucla.edu
\and
{\rm Yuval\ Tamir}\\
Computer Science Department, UCLA\\
tamir@cs.ucla.edu
}

\maketitle

\thispagestyle{empty}

\begin{abstract}
\name{} is a fully-operational fault tolerance mechanism
for multiprocessor workloads,
based on container replication,
using a hybrid of checkpointing and replay.
\name derives from two insights regarding replication mechanisms:
1)~deterministic replay can overcome a key disadvantage
of checkpointing alone --
unacceptably long delays of outputs to clients,
and
2)~checkpointing can overcome a key disadvantage
of active replication with deterministic replay alone --
vulnerability to even rare replay failures
due to an untracked nondeterministic events.
With \name,
the primary sends periodic checkpoints to the backup
and logs the outcomes of sources of nondeterminism.
Outputs to clients are delayed only by the
short time it takes to send the corresponding log
to the backup.
Upon primary failure,
the backup replays only the short interval
since the last checkpoint,
thus minimizing the window of vulnerability.
\name includes a ``best effort'' mechanism
that results in a high recovery rate even
in the presence of data races, as long as their rate is low.
The evaluation includes measurement of the recovery rate
and recovery latency based on fault injection.
On average, \name delays responses to clients by
less than 1ms and recovers in less than 1s.
For a set of eight real-world benchmarks, if
data races are eliminated, the performance
overhead of \name is under 59\%.
 \end{abstract}

\section{Introduction}
\label{sec:intro}

For many applications hosted in data centers,
high reliability is a key requirement,
demanding fault tolerance.
The fault tolerance techniques deployed are
preferably application-transparent
in order to avoid imposing extra burden on application
developers and facilitate use for legacy applications.
Replication, has long been used to implement
application-transparent fault tolerance, especially
for server applications.

The two main approaches to replication, specifically,
duplication, are:
(1)~high-frequency transfer of
the primary replica state (checkpoint), at the end of every execution
\textit{epoch}, to an inactive backup,
so that the backup can take over if the
primary fails~\cite{remus};
and
(2)~active replication, where the backup
mirrors the execution on the primary so that
it is ready to take over.
The second approach is challenging
for multiprocessor workloads, where there
are many sources of nodeterminism.
Hence, it is implemented in
a leader-follower setup, where the outcomes
of identified nondeterministic events, namely
synchronization operations
and certain system calls, on the primary are recorded
and sent to the backup, allowing the backup
to deterministically replay their outcomes~\cite{rex,castor}.

A key disadvantage of the first approach above is that,
for consistency between server applications
and their clients after failover,
outputs must be delayed and released
only after the checkpoint of the corresponding epoch
is committed at the backup.
Since checkpointing is an expensive operation,
for acceptable overhead,
the epoch duration is typically
set to tens of milliseconds.
Since, on average, outputs are delayed by half
an epoch, this results
in delays of tens of milliseconds.
A key disadvantage of the second approach is that it
is vulnerable to even rare replay failures
due to untracked nondeterministic events, such as those
caused by data races.

This paper presents a novel fault tolerance scheme,
based on container replication, called \name{}
(Hybrid Container Replication).
\name overcomes the two disadvantages above
using a unique combination of
periodic checkpointing~\cite{remus,nilicon},
externally-deterministic
replay~\cite{Chen15,doubleplay, respec, odr, pres},
user-level recording of nondeterministic events~\cite{respec,castor},
and failover of network connections~\cite{ft-tcp,coral}.
A critical feature of \name is that the checkpointing
epoch duration does not affect the response latency,
enabling \name to achieve sub-millisecond
added delay (\S\ref{subsec:eval_latency}).
This allows adjusting the epoch duration to trade off
performance and resource overheads with
recovery latency and
vulnerability to untracked nondeterministic events.
The latter is important since, especially legacy applications,
may contain data races (\S\ref{subsec:eval_recovery}).
\name is focused on dealing with data races
that rarely manifest and are thus more likely
to remain undetected.
Since \name only requires replay during recovery and
for the short interval since the last checkpoint, it is
inherently more reslilient to data races than
schemes that rely on replay of the entire execution~\cite{rex}.
Furthermore, \name includes a simple timing adjustment mechanism
that results in a high recovery rate even for applications
that include data races, as long as their rate of
unsynchronized writes is low.

Replication can be at the level of
VMs~\cite{Bressoud95, remus, phantasy, plover, colo},
processes~\cite{rex, castor},
or containers~\cite{nilicon}.
We believe that containers are the best choice
for mechanisms such as \name{}.
Applying \name{}'s approach to
VMs would be complicated since
there would be a need to track and replay
nondeterministic events in the kernel.
On the other hand, with processes, it is difficult
to avoid potential name conflicts upon failover.
A simple example is that the process ID used on
the primary may not be available on the backup.
While such name conflicts can be solved, the existing
container mechanism already solves them efficiently.

With \name{}, execution on the primary is divided into
epochs and the primary state is checkpointed to an inactive
backup at the end of each epoch~\cite{remus,nilicon}.
Upon failure of the primary, the backup begins execution
from the last primary checkpoint and then
deterministically replays the primary's execution
of its last partial epoch, up to the last external output.
The backup then proceeds with live execution.
To support the backup's deterministic replay,
\name ensures that, before an external output is released,
the backup has the log of nondeterministic events
on the primary since the last checkpoint.
Thus, external outputs are delayed only by the time
it takes to commit the relevant last portion of the
log to the backup.

Combining checkpointing with
externally-deterministic replay for replication
is not new~\cite{respec,rex,chenpatent,King05,castor}.
However, Respec~\cite{respec} requires an average external output
delay greater than half an epoch and is based on
\textit{active} replication.
\cite{chenpatent,King05}~do not provide support for execution
on multiprocessors.
See \S\ref{sec:related} for additional discussion.
Furthermore, these prior works do not provide
an evaluation of recovery rates
and are not designed or evaluated for containers.

We have implemented a prototype of \name and
evaluated its performance and reliability
using eight benchmarks.
We obtained the source code for NiLiCon~\cite{nilicon},
and it served as the basis for the implementation
of checkpointing and restore.
The rest of the implementation is new,
involving instrumenting standard library calls at the user level,
user-level agents, and small kernel modifications.
With 1s epochs,
\name's performance overhead was
less than 59\% for all eight benchmarks.
With more conservative 100ms epochs,
the overhead was less than 68\% for seven
of the benchmarks and 145\% for the eighth.
\name is designed to recover from fail-stop faults.
We used fault injection
to evaluate \name's recovery mechanism.
For all eight benchmarks, after data races
identified by ThreadSanitizer~\cite{threadsanitizer}
were resolved,
\name's recovery rate was 100\% for 100ms and 1s epochs.
Three of the benchmarks originally included data races.
For two of these, without any modifications,
with 100ms epochs and \name{}'s timing adjustments,
the recovery rate was over 99.4\%.

We make the following contributions:
1)~A novel fault tolerance scheme
based on container replication, using a unique
combination of
periodic checkpointing, deterministic replay
of multiprocessor workloads,
user-level recording of non-deterministic events,
and an optimized scheme for failover of network connections.
2)~A practical ``best effort'' mechanism that
enhances the success rate of deterministic replay
in the presence of data races
3)~A thorough evaluation of \name with respect
to performance overhead, resource overhead,
and recovery rate, demonstrating the lowest reported
external output delay compared to competitive mechanisms.

Section \ref{sec:background} presents two
key building blocks for \name: NiLiCon~\cite{nilicon}
and deterministic replay~\cite{Chen15,doubleplay, respec, odr, pres}.
An overview of \name is presented in \S\ref{sec:overview}.
\name{}'s implementation is described in \S\ref{sec:impl},
with a focus on key challenges.
The experimental setup and evaluation
are presented in \S\ref{sec:exp},
and \S\ref{sec:eval}, respectively.
Limitation of \name and of our prototype implementation
are described in \S\ref{sec:limit}.
\S\ref{sec:related} provides a brief overview of related work.
 \section{Background}
\label{sec:background}

\name{} integrates container replication
based on periodic checkpointing~\cite{remus,nilicon},
described in \S\ref{subsec:bg_nilicon},
and deterministic replay of multithreaded
applications on multiprocessors,
described in \S\ref{subsec:bg_drnr}.

\subsection{NiLiCon}
\label{subsec:bg_nilicon} 

Remus~\cite{remus} introduced a practical
application-transparent fault tolerance scheme
based on VM replication using high-frequency checkpointing.
NiLiCon~\cite{nilicon} is an implementation
of the Remus mechanism for containers.
With NiLiCon, the active primary container and
passive backup container are on different hosts.
Execution on the primary is divided into epochs.
At the end of each epoch, the primary is paused
and an incremental checkpoint, containing
all state changed since the last checkpoint,
is sent to the backup.
To track the memory state changes, NiLiCon
sets all the memory pages to be read-only at the beginning
of the epoch.
Thus, the first write to a page in an epoch
triggers a page fault exception, allowing
NiLiCon to record the page number and restore
its original access permissions.

As presented in \S\ref{sec:intro}, external outputs
(server replies to clients) are initially buffered
and are released only after the commitment on
the backup host of the checkpoint of the
epoch that generated the outputs.
If the primary fails, the checkpoint on the
backup host is used to restore the container.
Since no external outputs are released prior
to checkpoint commitment, consistency between
the containers and external clients is guaranteed,
even if the workload is non-deterministic.

NiLiCon's implementation is
based on a tool called CRIU (Checkpoint/Restore in User
Space)~\cite{criu} with
optimizations that reduce overhead.
CRIU checkpoints and restores the
user-level and kernel-level state of a container,
except for disk state.
NiLiCon handles disk state
by adding system calls to checkpoint and restore
the page cache and
a modified verion of the DRBD module~\cite{drbd_xen}.

To ensure consistency of the checkpointed state,
CRIU utilizes the Linux kernel's
\textit{freezer}~\cite{cgroup_freezer} feature
to stop the container so that its state does not
change during checkpointing.
A virtual signal to
all the threads in the container causes them to pause.
A thread executing a system call immediately exits the system call
and is paused before returning to user mode.
Once all the threads are paused, checkpointing proceeds.

NiLiCon relies on CRIU to preserve established TCP connections
across failover, using
a special repair mode of the
socket provided by the Linux kernel~\cite{tcp_repair}.
After setting a socket to repair mode,
a user-level process can
obtain or modify TCP state, such as sequence
numbers and buffered packets.

\subsection{Deterministic Replay on Multiprocessors}
\label{subsec:bg_drnr}

Deterministic replay is the reproduction
of some original execution in a subsequent execution.
During the original execution,
the results of
nondeterministic events/actions are recorded in a log.
This log is used in the subsequent execution~\cite{Chen15}.
With a uniprocessor, nondeterministic events
include: asynchronous events, such
as interrupts; system calls, such as gettimeofday;
and inputs from the external world.

With shared-memory multiprocessors,
there is a higher frequency of nondeterministic
events related to the order of accesses to
the same memory location by different processors.
Without hardware support, the cost of providing
deterministic replay of all such events is prohibitive.
Hence, a common practical approach is to support
deterministic replay only for programs
that are data-race-free~\cite{kendo}.
For such programs, as long as the results of synchronization
operations are deterministically replayed,
the ordering of shared memory accesses are preserved.
This involves much lower overhead since
the frequency of synchronization operations
is much lower than normal memory accesses.

The recording of nondeterministic events can
occur at different levels: hardware~\cite{fdr, rerun},
hypervisor~\cite{revirt, King05, smp-revirt}, OS~\cite{r2,
Laadan10}, or library~\cite{recplay, kendo}.
Without dedicated hardware support, the recording must be
done in software.
It is advantageous to record
the events at the user level, thus avoiding
the overhead for entering the kernel
or hypervisor~\cite{respec}.

The degree to which the replay must recover
the details of the original execution
depends on the use case~\cite{Chen15}.
To support seamless failover with replication,
it is sufficient to
provide \textit{externally deterministic replay}~\cite{respec}.
This means that, with respect to what is visible to external clients,
the replayed execution is identical
to the original execution.
Furthermore, the internal state at the
end of replay must be a state that corresponds
to a possible original execution that could result
in the same external behavior.
This latter requirement is needed so that the
replayed execution can transition to consistent live execution
at the end of the replay phase.
 \section{Overview of \name{}}
\label{sec:overview}

\begin{figure}[!t]
    \centering
    \includegraphics[width=\columnwidth]{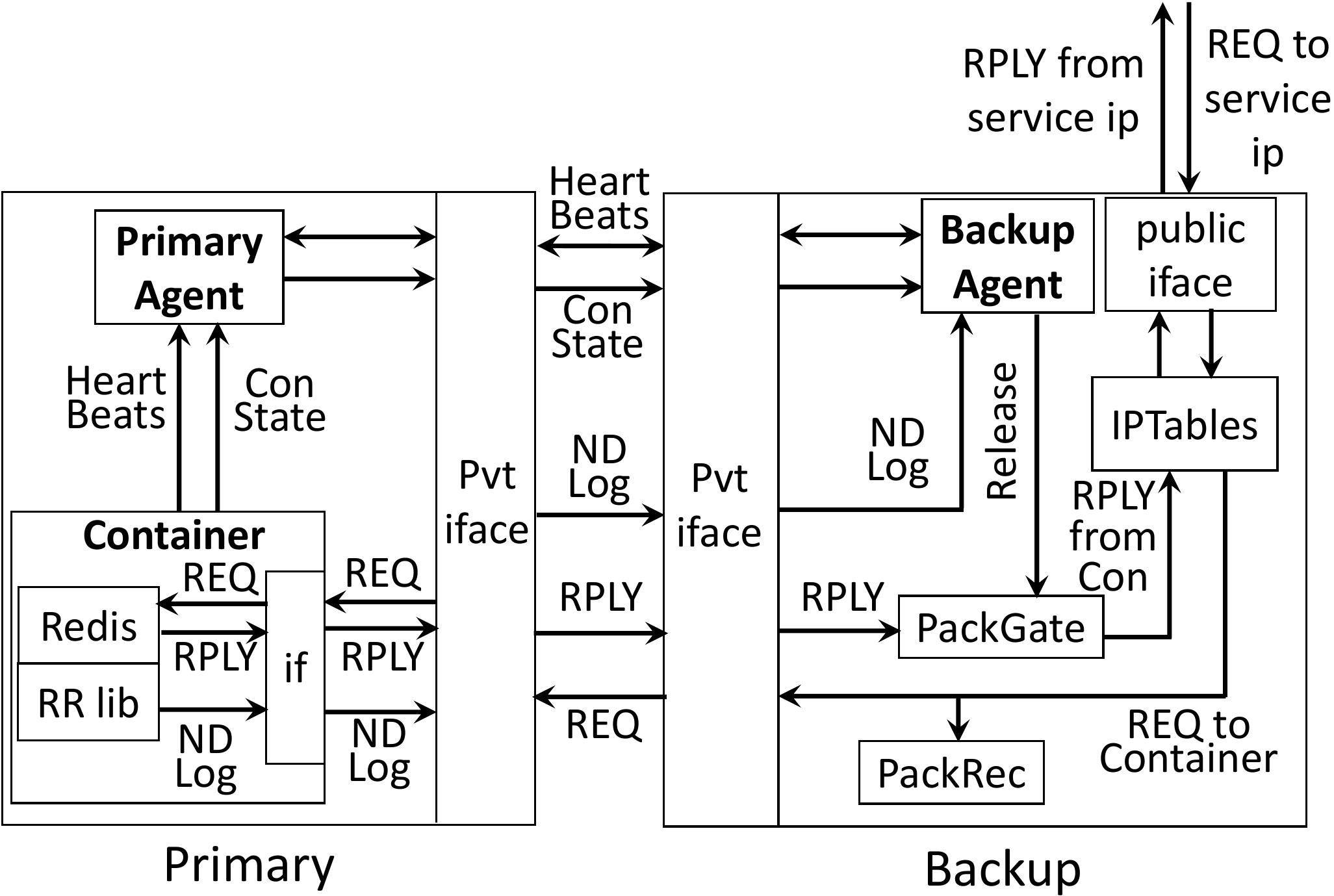}
    \caption{Architecture of \name{}.
             (ND log: non-deterministic event log).}
    \label{fig:arch}
\end{figure}

\name{} provides fault tolerance by maintaining
a primary-backup pair with an inactive backup
that takes over when the primary fails.
As discussed in
\S\ref{sec:intro}, this is done using a hybrid of
checkpointing and deterministic replay~\cite{chenpatent}.
The basic checkpointing mechanism uses the implementation that we
have obtained from the NiLiCon~\cite{nilicon} authors.

Figure~\ref{fig:arch} shows the overall architecture
of \name{}.
The primary records nondeterministic events:
operations on locks
and nondeterministic system calls.
The record and replay are done
at the user level, by instrumentation of glibc source
code.
When the primary executes,
the instrumented code invokes functions in
a dedicated RR (Record and Replay) library that
create logs used for replay.
There is a separate log for each lock.
For each thread, there is a log of the
nondeterministic system calls it invoked,
with their arguments and return values.
Details are presented in \S\ref{subsec:impl_rr}.

When the container sends a reply
to a client, the RR library collects the latest
entries (since the last transmission) of the nondeterministic
event logs and sends them to the backup.
To ensure consistency upon failover, the reply is
not released until the backup receives the
relevant logs.

\name does not guarantee recovery in the presence of
data races.
Specifically, unsynchronized accesses to
shared memory during the epoch in which the primary fails
may cause replay on the backup to fail to correctly
reproduce the primary's execution, leading the
backup to proactively terminate.
However, \name includes a simple ``best-effort''
mechanism that increases the probability of success
in such circumstances for application with a low rate
of unsynchronized accesses
to shared memory (\S\ref{subsec:impl_dataraces}).
With this mechanism, the order and timing of returns
from nondeterministic system calls by \textit{all}
the threads is recorded during execution on the primary.
During replay, the recorded order and relative
timing are enforced.

If the primary fails, network connections must
be maintained and migrated to the
backup~\cite{ft-tcp,ft-tcpj,coral,coralj}.
Like CoRAL~\cite{coral, coralj}, requests are routed through
backup by advertising the service IP address
in the backup.
Unlike FT-TCP~\cite{ft-tcp,ft-tcpj} or CoRAL,
replies are also routed through the backup,
resulting in lower latency (\S\ref{subsec:impl_nw}).

As with most other state replication work~\cite{remus, plover,
phantasy, nilicon}, \name{} assumes fail-stop faults.
Either the primary or the backup may fail.
Heartbeats are exchanged between the primary
and backup so failures are detected as missing heartbeats.
Handling of primary failures have already been discussed.
If the backup fails, the primary
configures its network, advertises the service IP address,
and communicates with the clients directly.
 \section{Implementation}
\label{sec:impl}

This section presents the implementation of \name{}, focusing
on the mechanisms used to overcome key challenges.
\name{} is implemented mostly at the user level
but also includes small modifications to the kernel.
At the user level, the implementation includes:
agent processes on the primary and backup hosts
that run outside the replicated container;
a special version of the glibc library (that includes Pthreads),
where some of the functions are instrumented (wrapped),
used by the application in the container;
and a dedicated RR (record and replay) library,
that provides functions that actually perform
the record and replay of nondeterministic events,
used by the application in the container.

The kernel modifications include:
an ability to record and enforce the order
of access to key data structures (\S\ref{subsec:impl_rr});
support for a few variables shared between the kernel
and RR library, used to coordinate checkpointing with record and
replay (\S\ref{subsec:impl_integrate});
and a new queueing discipline kernel module used to
pause and release network traffic (\S\ref{subsec:impl_nw}).

In the rest of this section,
\S\ref{subsec:impl_rr} presents the
basic record and replay scheme.
\S\ref{subsec:impl_integrate} deals with the challenge
of integrating checkpointing with record and replay.
\S\ref{subsec:impl_nw} presents the handling
of network traffic.
The transition from replay to live execution
is discussed in \S\ref{subsec:impl_recov}.
The performance-critical operation of transmitting
the nondeterministic event log to the backup
is explained in \S\ref{subsec:impl_optimization}.
\S\ref{subsec:impl_dataraces}
presents our best-effort mechanism for increasing
the probability of correct replay in
the presence of infrequently-manifested data races.

\subsection{Nondeterministic Events Record/Replay}
\label{subsec:impl_rr}

To minimize overhead and implementation complexity,
\name records synchronization operations and
system calls at the user level.
This is done by code added in glibc before
(\textit{before hook}) and after (\textit{after hook})
the original code.
Recording is done in the after hook,
replay is in the before hook.

For each lock there is a log of lock operations
in the order of returns from those operations.
The log entry includes the ID of the invoking thread
and the return value.
The return values is recorded to handle
the trylock variants as well as errors.
During replay, in most cases synchronization operations
must actually be performed in order to properly
enforce the correct semantics.
However, if the recorded return value indicates
that the thread failed to acquire the lock,
the thread directly returns instead of
trying to acquire the lock.
For each lock, the total ordering of returns
from each operation is enforced.
This is not really necessary for reader-writer
locks and trylock operations.
However, it simplifies the implementation and there
are minimal negative consequence since these events are
relatively rare.

For each thread, there is a log of invoked system calls.
The log entry includes the parameters and
return values.
During replay, the recorded parameters are used
to detect divergence (replay failure).
For some functions, such as gettimeofday(),
replay does not involve the
execution of the function and the recorded return
values are returned.
However, as discussed in \S\ref{subsec:impl_recov},
functions, such as open(), that involve the
manipulation of kernel state, are actually executed
during replay.

A key challenge is the replay of system
calls that are causally dependent.
These functions interact
with the kernel and \name does not
replay synchronization within the kernel~\cite{Laadan10}.
Thus, for example, if two threads invoke open()
at approximately the same time, without user-level
synchronization, the access within the kernel
to the file descriptor table may occur in a different
order during record and replay.
As a result, during replay, each thread would not obtain
the same file descriptor as it did during the original execution.

To meet the above challenge, \name uses a modified version
of the Rendezvous mechanism in Scribe~\cite{Laadan10}.
Specifically, the kernel is modified to maintain
an access sequence number for each shared kernel
resource, such as the file descriptor table.
Each thread registers
the address of a per-thread variable with the kernel.
When the
thread executes a system call accessing a shared resource,
the kernel increments the sequence number and copies its value
to the registered address.
At the user level, this sequence number is attached
to the corresponding system call log entry.
During replay, the before and after hooks enforces the recorded
execution order.

\subsection{Integrating Checkpointing with\\ Record/Replay}
\label{subsec:impl_integrate}

Two aspects of \name complicate the integration
of checkpointing with record/replay:
I)~the RR library and the data structures it maintains
are in the user level and are thus
part of the state that is checkpointed and 
restored with the rest of the container state;
and II)~checkpointing is triggered by a timer
in the agent, external to the container~\cite{nilicon},
and is thus not synchronized with the
recording of nondeterministic events on the primary.

Based on the implementation described so far, the above complications
can lead to the failure of \name in
two
key scenarios:
(1)~a checkpoint may be triggered while the RR library
is executing code that must not be executed in the replay mode, such
as sending the nondeterministic event log to the backup;
(2)~a checkpoint may be triggered while
a thread's execution falls between the beginning of
a before hook and the end of an after hook,
potentially resulting in a state from which replay
cannot properly proceed;

To handle Scenario (1), \name{} prevents the
checkpoint from occurring while any application thread is
executing RR library code.
Each thread registers with the kernel the
address of a per-thread \textit{in\_rr} variable.
In user mode, the RR library sets/clears
the \textit{in\_rr} when it respectively
enters/leaves the hook function.
An addition to the kernel code that handles the \textit{freezer}
virtual signal (\S\ref{subsec:bg_nilicon})
prevents the thread from being paused if
the thread's \textit{in\_rr} flag is set.
However, the virtual signal remains pending.
To prevent checkpointing from being unnecessarily delayed,
after checkpointing is requested by the agent,
threads are paused immediately
before entering or
after returning from RR library code.
A \textit{checkpointing} flag, shared between
the agent that controls checkpointing and the RR library code,
is used by the agent to indicate that checkpointing is requested,
causing the RR library code to invoke a
\textit{do nothing} system call,
thus allowing the virtual signal to pause the thread.

Scenario (2) cannot be handled as Scenario~(1) since
preventing checkpointing from occurring while a
thread is between the before hook and after hook could
delay checkpointing for a long time if the thread
is blocked on a system call, such as read().
To handle this problem, \name uses three variables:
two per-thread flags -- \textit{in\_hook}
and \textit{syscall\_skipped}, as well as a
global \textit{current\_phase} variable.
The addresses of these variables are registered with
the kernel and are accessed by kernel modifications
required by \name{}.
The \textit{current\_phase} variable is
in memory shared between the agent and the applications
in the container (the RR library code).
It indicates the current execution phase of the container
and is thus set to
\textit{record}, \textit{replay}, or \textit{live}.
In the record phase,
\textit{in\_hook} is set in the before hook and
cleared in the after hook.
Flag \textit{syscall\_skipped} is used
to indicate whether, during the record phase,
the checkpoint was taken before or after executing
the system call.
This flag is cleared in the before hook.
In kernel code executing a system call,
if \textit{current\_phase} is set to \textit{replay}
and \textit{in\_hook} is set,
the system call is skipped and
\textit{syscall\_skipped} is set.

Replay is performed in the before hook (\S\ref{subsec:impl_rr}).
During replay, if the after hook finds that \textit{in\_hook}
is set, that indicates that checkpointing occurred between
the before and after hooks.
Thus, if \textit{current\_phase} is \textit{replay} and
\textit{in\_hook} is set, the after hook passes control
back to the before hook.
This allows the system call to be correctly replayed.
For system calls that are actually executed
during replay (\S\ref{subsec:impl_rr}), there is
a need to determine whether the system call
was actually invoked during the record phase.
If it was, the system call must not be invoked again
during replay.
This required determination is accomplished based
on the \textit{syscall\_skipped} flag.

The key problem in Scenario (2) is relevant
for lock operations as well as for system calls.
The solution described above for system call
is thus also used, in a simplified form, for lock operations.
In this case, the \textit{syscall\_skipped} flag is
obviously not used.
In the after hook, if \textit{in\_hook} is found to be set,
the lock is released and control is passed to the before hook,
thus allowing enforcement of the order of lock acquires.

\subsection{Handling Network Traffic}
\label{subsec:impl_nw}

The current \name implementation assumes that
all network traffic is via TCP.
To ensure failure transparency with respect to clients,
there are three requirements that must be met:
(1)~client packets that have been acknowledged
must not be lost;
(2)~packets to the clients that have not been acknowledged
may need to be resent;
(3)~packets to the clients must not be released
until the backup is able to recover the primary state
past the point of sending those packets.

Requirements (1) and (2) have been handled
in connection with other mechanisms, such as
\cite{ft-tcp, ft-tcpj, coral, coralj}.
With \name{}, this is done by mapping the advertised
service IP address to the backup.
Incoming packets are routed through the backup, where they
are recorded by the PackRec thread in the agent,
using the \textit{libcap} library.
Outgoing packets are also routed through the backup.
To meet requirement~(2),
copies of the outgoing traffic are sent to the
backup as part of the nondeterministic event log.

The PackGate kernel module on the backup is used
to meet requirement~(3).
PackGate maintains a \textit{release sequence number}
for each TCP stream.
When the primary container sends an outgoing message,
the nondeterministic event log it sends to
the backup (\S\ref{sec:overview}) includes
a release request that updates the
stream's release sequence number.

PackGate operates frequently and must thus be efficient.
Hence, it is critical that it is
implemented in the kernel.
Furthermore, it must maintain fairness among the TCP streams.
These goals are met by maintaining a FIFO queue of release requests
that is scanned by PackGate.
Thus, PackGate avoids iterating through the streams
looking for packets to release and releases packets
based on the order of sends.

\subsection{Transition to Live Execution}
\label{subsec:impl_recov}

As with~\cite{Laadan10, rex} and unlike the deterministic
replay tools for debugging~\cite{flashback, jockey, respec,
doubleplay}, \name{} needs to transition from the replay mode to the
live mode.
The switch occurs when the backup replica finishes
replaying the nondeterministic event log, specifically,
when the last system call that generated an external
output during the original execution is replayed.
To identify this last call, after the checkpoint
is restored, the RR library scans the nondeterministic event log
and counts the number of system calls that generated
an external output.
Once replay starts, this count is decremented and
the transition to live execution is triggered when
the count reaches~0.

To support live execution, after replay, the kernel state must
be consistent with the state of the container and
with the state of the external world.
For most kernel state, this is achieved by actually executing
during replay system calls that change kernel state.
For example, this is done for system calls that
change the file descriptor table, such as open(),
or change the memory allocation, such as mmap().
However, this approach does not work
for system calls that interact with the external world.
Specifically, in the context of \name{}, these
are reads and writes on sockets associated
with a connection to an external client.
As discussed in \S\ref{subsec:impl_rr}, such calls are
replayed from the nondeterministic event log.
However, there is still a requirement of ensuring that,
before the transition to live execution,
the state of the socket, e.g., sequence numbers,
must be consistent with the state of the container
and with the state of external clients.

To overcome the above challenge,
when replaying system calls that affect socket state,
\name{} records the state changes on the sockets
based on the nondeterministic event logs.
When the replay phase completes,
\name{} updates all the sockets based on the recorded state.
Specifically, the relevant components of socket state are:
the last sent sequence number,
the last acknowledged (by the client) sequence number,
the last received (from the client) sequence number,
the receive queue, and the write queue.
The initial socket state is obtained from the checkpoint.
The updates to the sent sequence number and
the write queue contents are determined
based on writes and sends in the nondeterministic event log.
For the rest of the socket state, \name cannot rely
on the event log since some packets received and
acknowledged by the kernel may not have been read
by the application.
Instead, \name uses information
obtained from PackRec (\S\ref{subsec:impl_nw}).

With respect to incoming packets, once the container
transitions to live execution, \name must
provide to the container
all the packets that were acknowledged
by the primary but were not read by applications.
During normal operation, on the backup host, PackRec keeps copies
of incoming packets while PackGate extracts
the acknowledgment numbers on each outgoing stream.
If the primary fails, PackGate stops releasing
outgoing packets and it thus has the
last acknowledged sequence number of each incoming stream.
Before the container is restored
on the backup, PackRec copies the recorded incoming
packets to a log.
PackRec uses the information collected by PackGate
to determine when it has all the required (acknowledged)
incoming packets.
Using the information from the nondeterministic event log
and PackRec,
before the transition to live execution,
the packet repair mode (\S\ref{subsec:bg_nilicon}) is used to
restore the socket state so that it is consist
with the state of the container and the external world.

\subsection{Transferring the Event Logs}
\label{subsec:impl_optimization}

Whenever the container on the primary sends a message to an external
client, it must collect the corresponding
entries from the multiple nondeterministic
event logs (\S\ref{subsec:impl_rr})
and send them to the backup (\S\ref{sec:overview}).
Hence, the collection and sending of the log is
a frequent activity, which is thus performance critical.
To optimize performance, \name includes performance
optimizations, such as a specialized heap allocator
for the logs and maintaining a list of logs
that have been modified since the last time 
log entries were collected.
However, such optimizations proved to be insufficient.
Specifically, with one of our benchmarks, \textit{Memcached},
under saturation,
the performance overhead was approximately~300\%.

To address the performance challenge above,
\name{} offloads the transfer of the nodeterministic
event log from the application threads
to a dedicated \textit{logging thread} added
by the RR library to the application process.
With available CPU cycles, such as additional cores,
this minimizes interruptions in the
operation of the application threads.
Furthermore, if multiple application threads
generate external messages at approximately the
same time, the corresponding multiple transfers of the logs 
are batched together, further reducing the overhead.
When an application thread sends an external message,
it notifies the logging thread via a shared ring buffer.
The logging thread continuously
collects all the notifications in the ring buffer and then
collects and sends the nondeterministic logs to the backup.
To reduce CPU usage and enable more batching,
the logging thread sleeps for the minimum
time allowed by the kernel between scans of the buffer.

To minimize the performance overhead, \name
allows concurrent access to different logs.
Thus, one application thread may log a lock operation
concurrently with another application thread that is
logging a system call, while the logging thread is collecting
log entries from a third log for transfer to the backup.
This enables the logging thread to collect
entries from different logs out of execution order.
Thus, the collected log transferred to the backup
for a particular outgoing message may be missing log entries
on which some included log entries depend.
For example, for a particular application thread,
a log entry for a system call may be included but
the entry for a preceding lock operation may be missing.
This can result in an incomplete log,
leading to replay failure.

There are two key properties of \name that help address
the correctness challenge above:
A)~there is no need to replay the nondeterministic event
log beyond the last system call that outputs to
the external world, and
B)~when an application thread logs a system call that outputs to
the external world, all nondeterministic events on which 
this system call may depend are already logged
in nondeterministic event logs.
To exploit these properties,
the RR library maintains on the primary a global sequence number
that is accessible to the application threads and
the logging thread.
We'll refer to this sequence number as the
\textit{primary batch sequence number} (PBSN).
A corresponding sequence number is maintained on the backup,
which we'll refer to as \textit{backup batch sequence number} (BBSN).

When an application thread logs a system call that outputs to
the external world, it attaches the PBSN to the log entry.
When the logging thread receives a request to
collect and send the current event log, it increments
the PBSN before taking any other action.
Thus, any log entry corresponding to a system call
that outputs to the external world that is created
after the logging thread begins collecting the log,
has a higher sequence number.
When the backup receives the event log, it increments the BBSN.
If the primary fails, before replay is initiated on the backup,
all the nondeterministic event logs collected
during the current epoch are scanned and the
entries for system calls that output to the external world
are counted \textit{if} their attached sequence number
is not greater than the BBSN.
During replay, this count is decremented for each
such system call replayed.
When it reaches 0, relay terminates and live execution commences.

\subsection{Mitigating the Impact of Data Races}
\label{subsec:impl_dataraces}

Fundamentally, \name is based on being able to
identify all sources of non-determinism
that are potentially externally visible,
record their outcomes, and replay them when needed.
This implies that applications are expected
to be free of data races.
However, since \name only requires replay
of short intervals (up to one epoch),
it is inherently more tolerant to
rarely manifested data races than schemes that
rely on accurate replay of the entire execution~\cite{rex}.
As an addition to this inherent advantage of \name{},
this section describes an optional mechanism in \name
that significantly increases the probability of correct
recovery despite data races, as long as the manifestation
rate is low.

\name{} mitigates the impact of data races by
adjusting the relative timing of
the application threads during replay to 
approximately match the timing during the original execution.
As a first step, in the record phase,
the RR library records the order and the TSC (time stamp counter)
value when a thread leaves the after hook of a system call.
In the replay phase,
the RR library enforces the recorded order on threads before
they leave the after hook.
As a second step,
during replay, the RR library maintains
the TSC value corresponding to the time when the
after hook of the last-executed system call was exited.
When a thread is about to leave a system call after hook,
the RR library delays the thread until the difference between
the current TSC and the TSC of the last system call
is larger than the corresponding difference in the
original execution.
System calls are used as the basis for the timing adjustments
since they are replayed (not executed) and are thus
likely to cause the timing difference.
This mechanism is evaluated in \S\ref{subsec:eval_recovery}.
 \section{Experimental Setup}
\label{sec:exp}

All the experiments were hosted on Fedora 29 with
the 4.18.16 Linux kernel.
The containers were hosted using
runC~\cite{runc} (version 1.0.1), a popular container runtime
used in Docker.

Three hosts were used in the evaluation.
The primary and backup were hosted on 36-core
servers, using modern Xeon chips.
These hosts were connected to each other
through a dedicated 10Gb Ethernet link.
The clients were hosted on a 10-core server,
based on a similar Xeon chip.
The client host was in a different building,
interconnected through a Cisco switch, using 1Gb Ethernet.

Mechanisms like \name are most useful for server
applications.
The mechanism is stressed by applications
that manage significant state, execute frequent system calls
and synchronization operations, and
interact with clients at a high rate
through many TCP connections.
Hence, five of the benchmarks used were
in-memory databases handling short requests:
\textit{Redis}~\cite{redis},
\textit{Memcached}~\cite{memcached},
\textit{SSDB}~\cite{ssdb},
\textit{Tarantool}~\cite{tarantool}
and \textit{Aerospike}~\cite{aerospike}.
These benchmarks were evaluated
with 50\% read and 50\% write requests to 100,000 100B
records, driven by \textit{YCSB}~\cite{YCSB} clients.
The number of user
client threads ranged from 60 to 480.

The evaluation also included
a web server, \textit{Lighttpd}~\cite{lighttpd},
and two batch PARSEC~\cite{PARSEC} benchmarks:
\textit{swaptions} and \textit{streamcluster}.
\textit{Lighttpd} was evaluated using
20-40 clients retrieving a 1KB static page.
For \textit{Lighttpd},
benchmarking tools SIEGE~\cite{siege}, ab~\cite{ab}
and wget~\cite{wget} were used to evaluate, respectively,
the performance overhead, response latency, and recovery rate.
\textit{Swaptions} and \textit{streamclusters}
were evaluated using the native input test suites.

We used fault injection to evaluate
\name{}'s recovery mechanism.
Since fail-stop failures are assumed, a simple failure
detector was sufficient.
Failures were detected based on heart beats
exchanged every 30ms between the primary and backup hosts.
The side not receiving heart beats
for 90ms identified the failure of the other side
and initiates recovery.

For \textit{swaptions} and \textit{streamcluster}, recovery
was considered successful if the output was identical to
the golden copy.
For \textit{Lighttpd}, we used multiple
wget instances that concurrently fetched a static page.
Recovery was considered successful if all the fetched pages
were identical to the golden copy.
For the in-memory database benchmarks,
the \textit{YCSB} clients could not be used
since they do not verify the contents of the replies
and thus could not truly validate correct operation.
Instead, we developed customized clients, using
existing client libraries~\cite{hiredis, libmemcached, libtarantool,
libaerospike}, that spawns multiple threads and let each thread
work on separate set of database records.
Each thread records
the value it stores with each key, compares that value with
the value returned by the corresponding get operation and
flags an error if there is a mismatch.
Recovery was considered
successful if no errors were reported.

A possible concern with the customized client programs is that, due
to threads working on separate sets of database records,
lock contention is reduced and this could skew the results.
We compared the recovery
rate and recovery latency results of the customized clients
with the \textit{YCSB} clients.
For the \textit{YCSB} clients, recovery
was considered successful if replay succeeded and the clients
finished without reporting errors.
The results were similar:
the recovery rate difference was less than 2\%
and the recovery latency difference was less than 5\%.
In \S\ref{subsec:eval_recovery},
we report the more robust results obtained
with the customized client programs.

For the fault injection experiments,
for server programs, the clients
were configured to run for at least 30 seconds and drive the
server program to consume around 50\% of
the CPU cycles.
A fail stop failure was injected at a random
time within the middle 80\%
of the execution time, using the \textit{sch\_plug} module to
block network traffic on all the interfaces of a host.
To emulate a real world cloud computing environments, while
also stressing the recovery mechanism,
we used a \textit{perturb} program to
compete for CPU resources on the primary host.
The \textit{perturb} program
busy loops for a random time between 20 to 80 ms and sleeps
for a random time between 20 to 120ms.
During fault injection, a 
\textit{perturb} program instance was pinned
to each core executing the benchmark.
 \section{Evaluation}
\label{sec:eval}

This section presents \name{}'s performance overhead
and CPU usage overhead (\S\ref{subsec:eval_tp}),
the added latency for server
responses (\S\ref{subsec:eval_latency}), as well as the recovery
rate and recovery latency (\S\ref{subsec:eval_recovery}).
Two configurations of \name are evaluated:
\namese{} (short epoch) and \namele{} (long epoch),
with epoch durations of 100ms and 1s, respectively.
Setting the epoch duration is a tradeoff between
the lower overhead with long epochs and the
lower susceptibility to data races and lower
recovery time with short epochs.
Hence, \namele{} may be used if there is high confidence
that the applications are free of data races.
Thus, with the \namese{} configuration,
the data race mitigation mechanism described
in \S\ref{subsec:impl_dataraces} is turned on,
while it is turned off for \namele{}.

\name is compared to NiLiCon (\S\ref{subsec:bg_nilicon})
with respect to the
performance overhead under maximum CPU utilization and the
server response latency.
NiLiCon is configured to run with an
epoch interval of 30ms, as in~\cite{nilicon}.
The short epochs of NiLiCon are required since,
unlike \name{}, the epoch duration with NiLiCon
determines the added latency in replying
to client requests (\S\ref{subsec:bg_nilicon}).
In all cases, the ``stock setup'' is the
application running in an unreplicated container.

\subsection{Overheads: Performance, CPU Utilization}
\label{subsec:eval_tp}

Two measures of the overhead of \name are,
for a fixed amount of work, the
increase in execution time and the
increase in the utilization of CPU cycles.
These measures are distinct since
many of the actions of \name are in
parallel with the main computation threads.

For the six server benchmarks, the measurements
reported in this section were done with
workloads that resulted in maximum CPU utilization
for the cores running the application worker
threads\footnote{Some application
``helper threads'' are mostly blocked sleeping.}
with the stock setup.
To determine the required workloads, the number of client
threads was increased until an increase
by 20\% resulted in an increase of less than 2\%
in throughput.
This led to CPU utilization of above 97\% for
all the worker threads except with \textit{SSDB}.
With \textit{SSDB} a bottleneck thread 
resulted in utilization of 98\%, while the rest
resulted in utilization of approximately 48\%.
Additional measurements were done to verify the
network bandwidth was not the bottleneck.

With four of the server benchmarks, the number of the
worker threads cannot be configured
(\textit{Lighttpd}, \textit{Redis}: 1,
\textit{Tarantool}: 2, \textit{SSDB}: 12).
\textit{Mecached}, \textit{Aerospike} were configured
to run with four worker threads.
For these applications, the number of the worker threads
is set to four because, with our experimental setup,
it was not possible to generate enough client traffic
from YCSB to saturate more than four worker threads.
For consistency, the two non-interactive benchmarks
were also configured to run with four threads.

For each benchmark, the workload that saturates
the cores in the stock setup was used for
the stock, \name{}, and  NiLiCon setups.
With NiLiCon,
due to the long latencies it normally incurs for server responses
(\S\ref{subsec:eval_latency}),
it is impossible to saturate the server with this setup.
Hence, for the NiLiCon measurements in this subsection, the
buffering of the server responses was removed.
This is not a valid NiLiCon configuration, but it
provides a comparison of the overheads excluding buffering
of external outputs.

\begin{figure}[!t]
    \centering
    \includegraphics[width=\columnwidth]{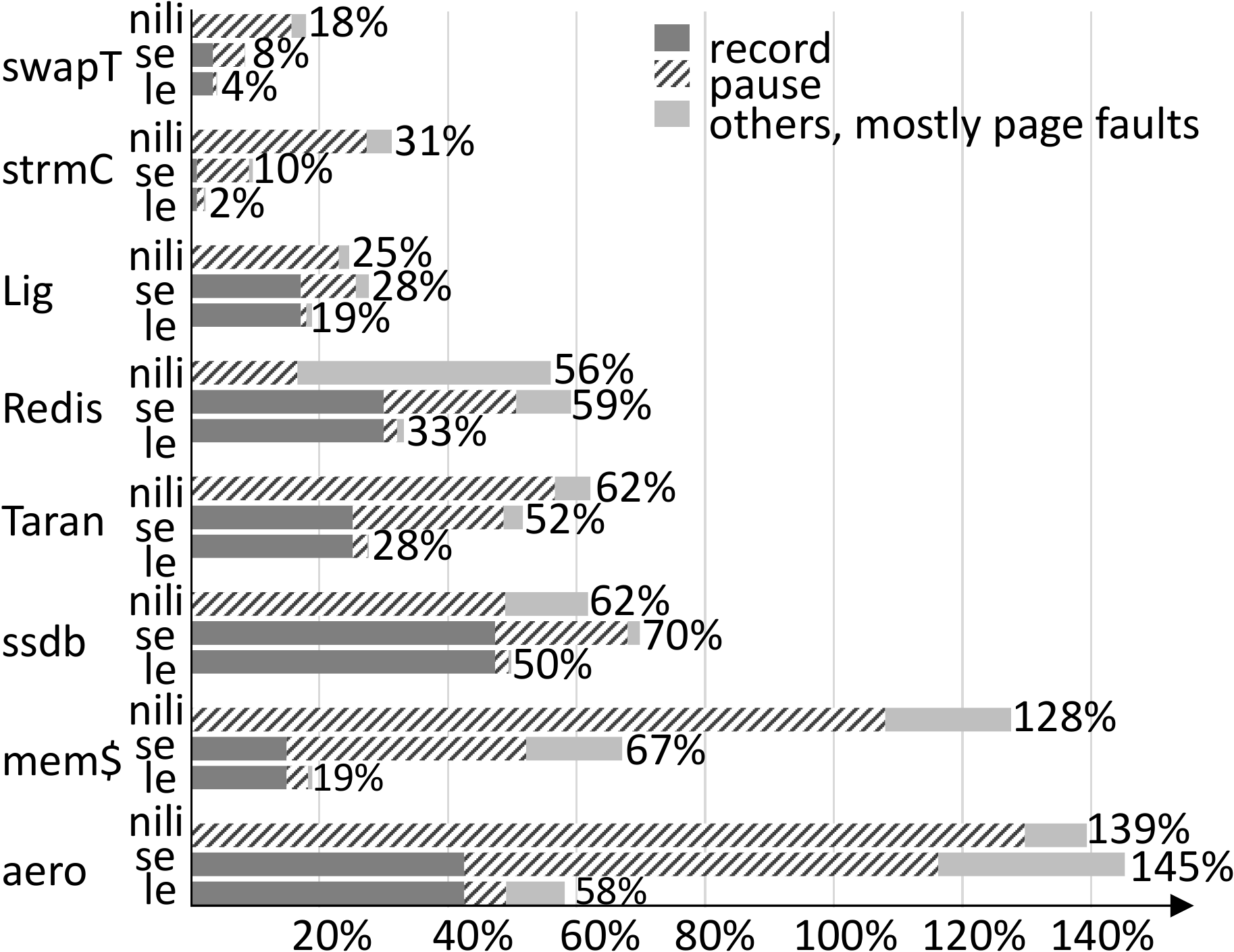}
    \caption{Performance overheads: NiLiCon, \namese{},
    \namele{}.}
    \label{fig:tp}
\end{figure}

\textbf{Performance Overhead.}
The performance overhead
is reported as the percentage increase in the execution
time for a fixed amount of work
compared to the stock setup.
Figure~\ref{fig:tp} shows the performance overheads
of NiLiCon, \namese{}, and \namele{}, with the breakdown of
the sources of overhead.
Each benchmark was executed 50 times.
The margin of error of the 95\% confidence
interval was less than 2\%.

The record overhead is caused by the RR library recording
non-deterministic events.
The pause overhead is due to the time
the container is paused for checkpointing.
The page fault
overhead is caused by the page fault exceptions that track the
memory state changes of each epoch (\S\ref{subsec:bg_nilicon}).

As shown by a comparison of the results for \namese{} and \namele{},
due to locality in accesses to pages,
the pause and page fault overheads decrease as
the epoch duration is increased.
This comparison also shows that the fact that the
data race mitigation mechanism is on with \namese{} and of
with \namele{}, has no significant impact on the record overhead.
With \namese{},
the average incremental checkpoint size per epoch was
0.2MB for \textit{Swaptions},
15.6MB for \textit{Redis},
and
41.2MB for \textit{Aerospike},
partially explaining the differences in pause overhead,
which is also affected by the time to obtain
required kernel state~\cite{nilicon}.
With \namese{},
the average number of logged lock operations plus system calls
per epoch was
9 with \textit{streamcluster},
907 with \textit{Tarantool},
and
2137 with \textit{Aerospike},
partially explaining the differences in record overhead.
However, the overhead of logging system calls is much
higher than for lock operations.
\textit{Memcached} is comparable to \textit{Aerospike}
in terms of the rate of logged system calls plus lock operations,
but has 341 compared to 881 logged system calls
per epoch and thus lower record overhead.

\textbf{CPU utilization overhead.}
The CPU utilization is the product of the average numbers
of CPUs (cores) used and the total execution time.
The CPU utilization overhead is the percentage increase
in utilization with \name compared to with the stock setup.
The measurement is done by pinning each \name{}
component to a dedicated set of cores.
All user threads are configured to run at high priority
using the SCHED\_FIFO real-type scheduling policy.
Instances of a simple program
that continuously increments a
counter run at low priority on the different cores.
The CPU utilization is determined by comparing the
values of counts from those instances to the
values obtained over the same period on an idle core.
The experiment is repeated 50 times, resulting
in a 95\% confidence interval
margin of error of less than~1\%.

\begin{table}
    \centering
    \newcommand{\rot}[2]{\multicolumn{1}{#1}{\rotatebox{90}{#2}}}
\newcommand{\txt}[1]{\multicolumn{1}{c|}{#1}}
\newcommand{\ltxt}[1]{\multicolumn{1}{c}{#1}}

\renewcommand{\arraystretch}{1.2}
\setlength{\tabcolsep}{0.4ex}

\begin{tabular}
    { cr V{3.0}*{7}{r|}
      r }
      & & \txt{ST} & \txt{SC} & \txt{Lig} & \txt{Redis} & \txt{Taran} & \txt{SSDB} & \txt{Mem\$} & \ltxt{Aero}
\\

\hlineB{2.5}

\multirow{2}{*}{\textbf{P}} & LogTH   &  $\sim$0 & $\sim$0  & 17\%  & 11\%  & 12\%  & 5\% & 12\% & 20\% 
\\\cline{2-10}
                            & others   &     6\%  & 3\%  & 27\%  & 63\%  & 45\% & 55\% & 36\% & 87\%
\\\hlineB{2.5}
\multirow{3}{*}{\textbf{B}} & KerNet &  $\sim$0  & $\sim$0 & 43\%  & 70\%  & 43\%  & 18\%  & 31\% & 45\%
\\\cline{2-10}
                            & PKRec    &  $\sim$0  & $\sim$0 & 17\% &  15\% & 10\% &  4\%  &  9\% & 13\%
\\\cline{2-10}
                            & others   &  1\% & 2\%  & 41\% & 54\% & 35\% & 15\% & 13\% & 20\%
\\\hlineB{2.5}
                            & total    &  7\% & 5\%  & 145\% & 213\% & 145\% & 97\% & 101\% & 185\%
\\\hline

\end{tabular}

     \caption{CPU utilization overhead for \namese{}.
             LogTH: logging thread. KerNet: kernel's
             handling of network packets. P: primary
	     host and B: backup host.}
    \label{table:cpu}
\end{table}

Table~\ref{table:cpu}
shows a breakdown of CPU utilization overhead with \namese{}.
The ``others'' row for the primary is the overhead
for recording the nondeterministic events,
handling the page fault exceptions for tracking memory changes,
and collecting and sending the incremental checkpoints.
The ``others'' row for the backup is the overhead
for receiving and storing the nondeterministic event logs
and the checkpoints.
The KerNet row for the backup is the overhead for packet
handling in the kernel, that includes the routing
of requests and responses to/from the primary
and the PackGate module.

In terms of CPU utilization overhead, the worst
case is with \textit{Redis}.
A significant factor is the overhead for packet
handling in the backup kernel (KerNet).
We have found that this overhead is
mostly due to routing, not PackGate.
\textit{Redis} involves only one worker thread and it
receives and sends
a large number of small packets, leading to this overhead.
Techniques for optimizing software
routing~\cite{click} can be used to reduce this overhead.

With \namele{}, the CPU utilization overhead
is 2\% to 169\%, with
\textit{Redis} still being the worst case.
The CPU utilization on
the primary is 1\% to 69\% -- significantly less
than that with \namese{} due to the reduction in CPU time to
handle checkpointing and page faults.
The CPU utilization overhead on the backup
is only slightly lower than with \namese{},
due to the reduction in CPU time to receive checkpoints.

\subsection{Response Latency}
\label{subsec:eval_latency}

\begin{table}
    \centering
    \newcommand{\txt}[1]{\multicolumn{1}{c|}{#1}}
\newcommand{\ltxt}[1]{\multicolumn{1}{c}{#1}}

\renewcommand{\arraystretch}{1.2}
\setlength{\tabcolsep}{0.4ex}

\begin{tabular}
    { cr V{3.0}*{6}{r|}
      r }
      & & \txt{Lig1K} & \txt{Lig100K} & \txt{Redis} & \txt{Taran} & \txt{SSDB} & \txt{Mem\$} & \ltxt{Aero}

\\

\hlineB{2.5}

\multirow{2}{*}{\textbf{S}} & avg & 549 & 2059    & 406  & 393  & 388  & 643 & 373
\\\cline{2-9}
                            & 99\% & <1ms & <3ms  & 734  & 617  & 622  & 2982 & 711
\\\hlineB{2.5}
\multirow{2}{*}{\textbf{H}} & avg  & 740  & 2215  & 637  & 651  & 709  & 1092 & 945
\\\cline{2-9}
                            & 99\% & <1ms & <9ms  & 1105 & 1191 & 1087 & 5901 & 1724
\\\hlineB{2.5}

\multirow{2}{*}{\textbf{N}} & avg  & 38ms  & 38ms  & 42ms & 42ms & 45ms & 45ms & 51ms
\\\cline{2-9}
                            & 99\% & <39ms & <39ms & 44ms & 42ms & 47ms & 53ms & 63ms

\\\hline

\end{tabular}

     \caption{Response Latency in $\mu$s. S: Stock, H: \namese{},
             N: NiLiCon}
    \label{table:response-latency}
\end{table}

A key advantage of \name{}
compared to schemes based on checkpointing alone,
such as Remus~\cite{remus} and NiLiCon~\cite{nilicon}
is significantly lower response latency.
Table~\ref{table:response-latency} shows the response
latencies with the stock setup, \namese{}
and NiLiCon.
The number of client threads for stock and \namese{}
is separately adjusted so that the CPU load on the
cores running application worker threads is
approximately 50\%.
For NiLiCon, due to
its long response latencies, it is not possible to reach
50\% CPU usage.
Instead, NiLiCon is evaluated with the same number
of client threads as \namese{}, resulting
in CPU utilization of less than 5\%, thus
favoring NiLiCon.
To evaluate the impact of response size,
\textit{Lighttpd} is evaluated serving both
1KB as well as 100KB files.
Each benchmark is executed 50 times.
We report the average of the mean and
the 99th percentile latencies of the different runs.
For the average response latencies,
the 95\% confidence interval has a margin
of error of less than 5\%.
For the 99th percentile latencies, it is less than 15\%.

With \name, there are three potential sources for 
the increase in response latency:
forwarding packets through the backup,
the need to delay packet release until the
corresponding event log is received by the backup,
and increased request processing time on the primary.
With \namese{}, the increase in average latency
is only 156$\mu$s to 581$\mu$s.
The worst case is with \textit{Aerospike},
which has the highest processing overhead (Fig.~\ref{fig:tp})
and a high rate of nondeterministic events and thus
long logs that have to be transferred to the backup.
The increase in 99th percentile latency is
371$\mu$s to 6ms.
The worst case is with \textit{Lighttpd} serving a 100KB file.
This is because the request service time is much longer
than with the other benchmarks and thus
a checkpoint is more likely to happen in the middle
of this time.
The pause time for checkpoint of this benchmark is
approximately 6ms.
It should be noted that, in terms of increase in
response latency, NiLiCon is not competitive,
as also indicated by the results in~\cite{nilicon}.

With \namele{}, the increase in the average response latency
is from 40$\mu$s to only 343$\mu$s, due to the
the lower processing overhead.
The increase in the 99th
percentile latency is under 534$\mu$s since
checkpoint are much less frequent and thus less likely
to interrupt the processing of a request.

\subsection{Recovery Rate and Latency}
\label{subsec:eval_recovery}

This subsection presents an evaluation of
the recovery mechanism and the data race mitigation mechanism.
The service interruption time is obtained
by measuring, at the client,
the increase in response latency when a fault occurs.
The service
interruption time is the sum of the recovery latency plus the
detection time.
With \name{}, the average detection time is
90ms (\S\ref{sec:exp}).
Hence, since our focus is not on detection mechanisms,
the average recovery latency
reported is the average service interruption time minus 90ms.

\textbf{Backup Failure.}
50 fault injection runs are
performed for each benchmark.
Recovery is always successful.
The service interruption duration is dominated by
by the Linux TCP retransmission timeout, which is 200ms.
The other recovery events, such as detector
timeout and broadcasting the ARP requests to update the service
IP address, occur concurrently with this 200ms.
Thus, the measured service interruption duration is
between 203ms and 208ms.
The 95\% confidence interval margin of error is
less than 0.1\%.

\textbf{Primary Failure Recovery Rate.}
Three of our benchmarks contain data races
that may cause recovery failure:
\textit{Memcached}, \textit{Aerospike} and
\textit{Tarantool}.
Running \textit{Tarantool}
with \namese{}, through 50 runs of fault injection
in the primary, we find that, due to data races,
in all cases replay fails and thus recovery fails.
Due to the high rate of data race manifestation,
this is the case even with the mechanism
described in~\S\ref{subsec:impl_dataraces}.
Thus, we use a modified version of \textit{Tarantool}
in which the data races are eliminated by manually
adding locks.

We divide the benchmarks into two sets.
The first set consists of the five data-race-free
benchmarks and a modified version of \textit{Tarantool}.
For these, 50 fault injections are
performed for each benchmark.
Recovery is always successful.

\begin{table}
    \centering
    \newcommand{\rot}[2]{\multicolumn{1}{#1}{\rotatebox{90}{#2}}}
\newcommand{\txt}[1]{\multicolumn{1}{c|}{#1}}
\newcommand{\stxt}[1]{\multicolumn{1}{cV{3.0}}{#1}}
\newcommand{\ltxt}[1]{\multicolumn{1}{c}{#1}}

\renewcommand{\arraystretch}{1.2}
\setlength{\tabcolsep}{0.4ex}

\begin{tabular}
    {l l V{3.0} r | r V{3.0} r|
      r }
      & & \multicolumn{2}{cV{3.0}}{Recovery Rate} & \multicolumn{2}{c}{Replay Time}
      \\\cline{3-6}
      & & \txt{Mem\$} & \stxt{Aero} & \txt{Mem\$} & \ltxt{Aero} 
    \\\hlineB{2.5}
    \multirow{3}{*}{\rotatebox{90}{100ms}}   & stock & 94.1\% & 83.4\% & 23 &  33
    \\\cline{2-6}
    & + Total order of syscalls &  93.9\% & 92.8\% & 128 & 289
    \\\cline{2-6}
    & + Timing adjustment & 99.5\% & 99.8\% & 234 & 377
    \\\hlineB{2.5}
    \multirow{3}{*}{\rotatebox{90}{1s}}   & stock & 50.2\% & 35.3\% & 245 & 370
    \\\cline{2-6}
    & + Total order of syscalls &  50.6\% & 78.1\% & 1129 & 1342
    \\\cline{2-6}
    & + Timing adjustment & 98.7\% & 99.2\% & 1218 & 1474

\end{tabular}

     \caption{Recovery rate and replay time (in ms).
             \name{} with different levels of
              mitigation of data race impact.}
    \label{table:datarace}
\end{table}

The second set of benchmarks consists of \textit{Memcached}
and \textit{Aerospike}, used to evaluate the
the data race mitigation mechanisms (\S\ref{subsec:impl_dataraces}).
For these, to ensure statistically significant results,
1000 fault injection runs
are performed with each benchmark with each setup.
The results are presented in Table~\ref{table:datarace}.
For both the recovery rate and replay time,
the 95\% confidence interval is less than~1\%.
Without the \S\ref{subsec:impl_dataraces} mechanism,
the recovery rate for \namele{} is much lower
than with \namese{}, demonstrating the benefit
of short epochs and thus shorter replay times.
Enforcing a total order of the recorded system calls
in the after hook is not effective for \textit{Memcached}
but increases the recovery rate of \textit{Aerospike}
for both \name setups.
However, with the timing adjustments, both benchmarks achieve
high recovery rates, even with \namele{}.
The total order of the system calls
is the main factor that increase the replay time.
Thus, there is no reason to not also enable
the timing adjustments.

To explain the results above, we measured the rate of
racy memory accesses in \textit{Tarantool}, \textit{Memcached}
and \textit{Aerospike}.
To identify ``racy memory accesses'', we first
fixed all the identified data races by protecting
certain memory access with locks.
We then removed the added locks and added instrumentation
to count the corresponding memory accesses.
For \textit{Tarantool}, the rates of racy
memory writes and reads are, respectively,
328,000 and 274,000 per second.
For \textit{Memcached} the respective rates are 1 and 131,000
per second and for \textit{Aerospike} they are
250 and 372,000 per second.
These results demonstrate that when the rate of
accesses potentially affected by data races is high
our mitigation scheme is not effective.
Fortunately, in such cases, data races are
unlikely to remain undetected.

As an additional validation of \name,
the three benchmarks mentioned above were 
modified to eliminate the data races.
With the \namele{} setup,
200 fault injection runs are executed
with \textit{Memcached} and \textit{Aerospike}.
50 fault injection runs are executed
with the remaining six benchmarks.
Recovery is successful in all cases.

\begin{figure}[!t]
    \centering
    \includegraphics[width=\columnwidth]{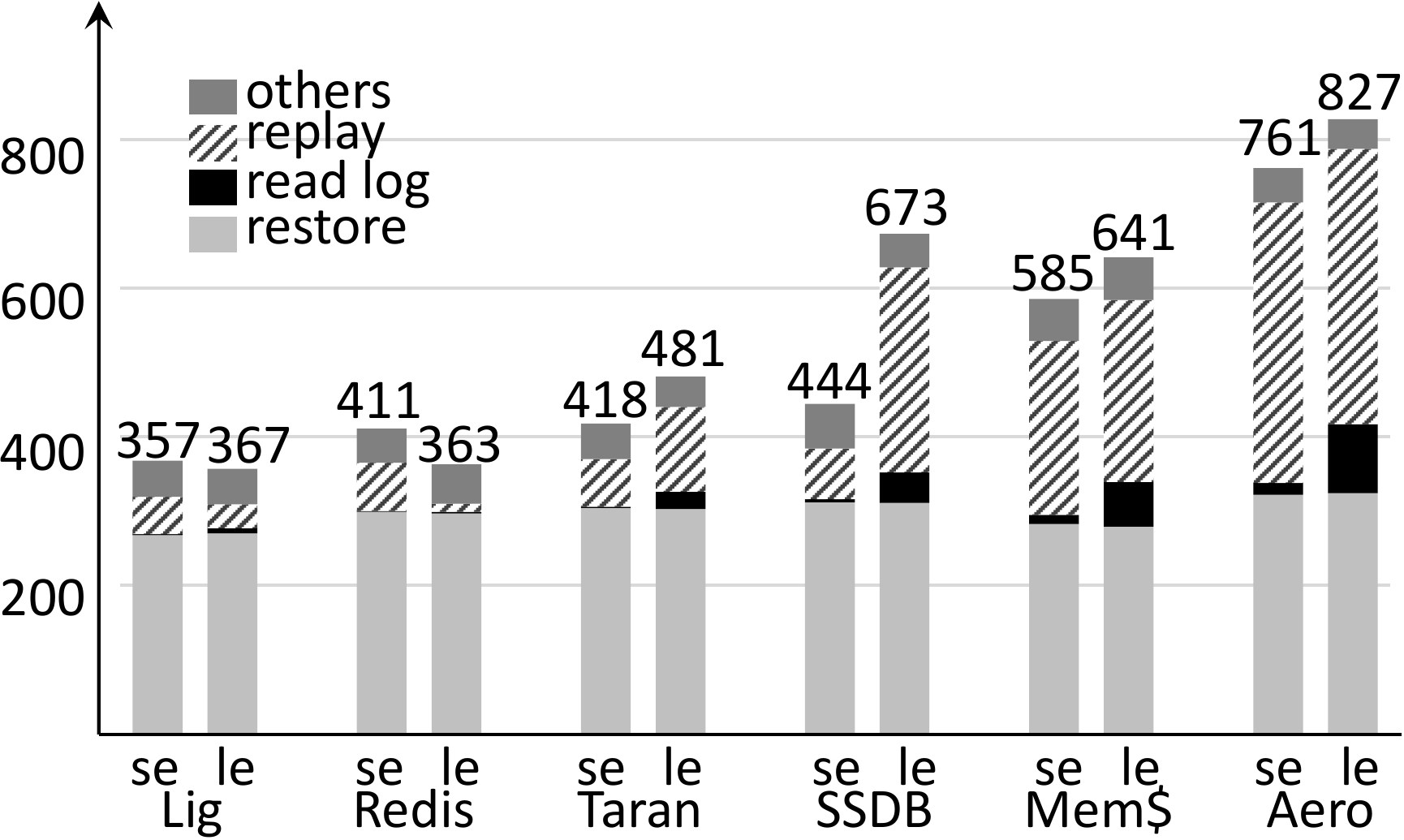}
    \caption{Recovery Latency (ms) breakdown with \namese{}
             and \namele{}.}
    \label{fig:recovery}
\end{figure}

\textbf{Primary Failure Recovery Latency.}
Figure~\ref{fig:recovery} shows a breakdown of
the factors that make up the recovery latency with
\namese{} and \namele{}.
The batch benchmarks, \textit{swaptions} and \textit{streamcluster},
are not included since their execution times are
above 60s, so their recovery latency is insignificant.
With \namese{}, the data race mitigation scheme is
enabled, while with \namele{} it is disabled.
The 95\% confidence interval margin of error is less than~5\%.
\textit{Restore} is the time to restore the checkpoint,
mostly for restoring the in-kernel states of the container
(e.g., mount points and namespaces).
\textit{Read log} is the time to process the stored logs
in preparation for replay.
\textit{Others} include the time to send ARP requests and connect
the backup container network interface to the bridge.

The recovery latency differences among the benchmarks are
due mainly to the replay time.
It might be expected that the average replay time would
be approximately half an epoch duration.
However, replay time is increased due to different
thread scheduling by the kernel that causes
some threads to wait to match the order of
the original execution.
This increase is more likely when the data race impact
mitigation mechanism is enabled since it enforces
more strict adherence to the original execution.
A second factor that impact the replay time
is a decrease due to
system calls that are replayed from the log and not executed.
 \section{Limitations}
\label{sec:limit}

We have identified one inherent limitation of \name
and four limitations of the current research
prototype implementation.
An inherent limitation is that the mechanism
used for mitigating the impact of
data races (\S\ref{subsec:impl_dataraces})
is incapable of handling a high rate
of racy accesses (\S\ref{subsec:eval_recovery}).
However, as discussed in \S\ref{subsec:eval_recovery},
such data races are easily detectable and are thus
easy to eliminate, even in legacy applications.

\name{} does not currently support
multiple processes.
To overcome this limitation, the RR library
would need significant enhancements, such
as support for inter-process communications via shared memory.
Techniques presented in~\cite{dos} may be applicable.
\name also does not handle asynchronous signals.
This may be resolved by techniques
used in~\cite{Laadan10}, that delay signal delivery until a
system call or certain page faults.

\name{} does not handle C atomic types, functions,
intrinsics and inline assembly code that performs atomic operations
transparently.
In this work, such cases were handled by protecting
such operations with locks.
Specifically, this was done for \textit{Aerospike} and
\textit{glibc}.
Compiler support~\cite{castor} is needed to overcome this limitation.

Recovery currently fails if a socket
is created via accept() or connect() during replay.
Resolving this limitation would require recording
and restoring during replay various socket state components,
such as the timestamp and window scale.
This can be done by enhancing the RR library
with code used in NiLiCon~\cite{nilicon}
to checkpoint and restore socket state.
 
\section{Related Work}
\label{sec:related}

\name{} builds on prior works on
fault tolerance using replication,
replication based on high-frequency checkpointing,
replication based on deterministic replay,
and network connection failover.
In the works cited, a few are
validated using fault
injection~\cite{remus,colo,coralj,Tardigrade,nilicon}.

As with \name{}, there are other replay systems that support
replay from a checkpointed
state~\cite{flashback,jockey,respec,Laadan10}
and the transition from replay to
live execution~\cite{Laadan10,rex}.
There are various works aimed at replay in the
presence of data races while only
recording system calls and synchronization
operations~\cite{odr, pres, respec, doubleplay}.
However, these
works either require delaying outputs until the
end of a long (tens of ms) epoch~\cite{respec, doubleplay},
or requires lengthy offline processing to enumerate the order
of memory accesses involved in the
data races~\cite{odr,pres},
making them unsuitable for fault tolerance.

Early work on VM replication for fault tolerance is based on
leader-follower active replication using
deterministic replay~\cite{Bressoud95}.
This is combined with periodic
checkpointing in~\cite{chenpatent}, based on
earlier work on using these technique for debugging~\cite{King05}.
Both of these works are focused on uniprocessor systems.
Extending them to multiprocessors
is impractical, due to the overhead of recording shared
memory access order in the
hypervisor~\cite{smp-revirt, samsara}.
Remus~\cite{remus} focuses on multiprocessors and implements
VM replication for fault tolerance using
high-frequency checkpointing alone (\S\ref{subsec:bg_nilicon}).
Tardigrade~\cite{Tardigrade} applies Remus's algorithm
to a lightweight VMs based on a library OS.
NiLiCon~\cite{nilicon} applies Remus's algorithm to containers.
Phantasy~\cite{phantasy} uses hardware features, PML
and RDMA, to optimize Remus.
Plover~\cite{plover} optimizes Remus
by using an active replica to reduce the size of transferred state
and by performing state synchronization adaptively,
when VMs are idle.
All the Remus-based mechanisms release outputs only
after the primary and backup synchronize their states,
Hence, outputs are delayed by multiple (often, tens of) milliseconds.
COLO~\cite{colo} uses active VM replication,
comparing outputs before release to the client.
On a mismatch, the state of one VM is updated with the other's.
There is no mechanism to ensure that the backup's
execution matches the primary's, resulting in
high performance overhead and long response latencies
for applications with significant nondeterminism.

Another set of works use deterministic replay,
discussed in~\S\ref{subsec:bg_drnr},
for active replication of multiprocessor
workloads~\cite{respec,rex,castor}.
As with \name{},
the primary records the outcomes of nondeterministic events
and logs them to the backup.
Rex~\cite{rex} requires the application to be data race free
and requires manual modifications of the application source
code to use a specified API.
As with \name{}, output to the external world
can be released only after the backup receives
the non-deterministic log.
Execution divergence, due to a data race or some other
unlogged nondeterministic event, can cause failure.
Castor~\cite{castor} handles data races by buffering the
output to the external world until the backup finishes replaying
the associated log.
If divergence, due to a data race,
prevents the backup from continuing replay,
the backup's state is synchronized with the primary's.

Comparing \name with Rex and Castor, the key difference
is the use of checkpointing versus an active replica.
The disadvantages of \name are
periodic pauses for checkpointing during normal operation
and higher recovery latency.
To explain the advantages of \name{}, two
cases are considered:
(1)~the applications are assumed to be free of data races, and
(2)~there may be some data races.
For the former case, it should be noted that,
for applications with a large number of
synchronization operations, replay may be slower
than the original execution.
Thus, under heavy load,
the active replica is a performance bottleneck~\cite{rex}.
For example, both Rex and \name are evaluated
with \textit{Memcached}, and the performance overheads
are 40\% and 19\%, respectively.

For applications that have data races, the only relevant
comparison is with Castor.
Castor is likely to have higher response delays
since outputs cannot be released until
the backup finishes replaying the associated log.
Additionally, with Castor as with \name{},
a data race can cause recovery to fail.
Specifically, with Castor, if the primary fails while transferring
its state to the backup, the system fails.
Hence, for an application with a high rate of
racy memory accesses, such as
\textit{Tarantool} (\S\ref{subsec:eval_recovery}),
Castor would be frequently synchronizing the backup state
and thus have low recovery rate (like \name{})
and also high performance overhead.
For applications with a lower rate of
racy memory accesses, such as
\textit{Memcached} and \textit{Aerospike},
execution divergence is less likely.
Whether \name or Castor have higher recovery rate
depends on the rate of execution divergence
and the cost of synchronizing the state.
For example, based on the recovery rate
for \textit{Memcached} shown in Table~\ref{table:datarace}
for the ``stock'' \namese{}, the probability
of execution divergence in 50ms (half an epoch) is 0.059.
Hence, execution diverges approximately every 0.85s.
With our setup, the time it takes to create
and transfer the checkpoint for \textit{Memcached}
is 48ms.
Hence, an upper bound on the
recovery rate with Castor is expected be 94.7\%
versus 99.5\% with \name (Table~\ref{table:datarace}).
A similar calculation for \textit{Aerospike},
taking into account 76ms to
create and transfer the checkpoint, results
in a recovery rate for Castor of 79.8\% versus
99.8\% for \name.
For programs with higher memory working set, the
state transfer time would be larger and thus the
recovery rate advantage of \namese{} would also be larger.

 \section{Conclusion}
\label{sec:concl}

\name is a unique point in the design space of
application-transparent fault tolerance schemes
for multiprocessor workloads.
By combining checkpointing with externally deterministic
replay, it facilitates trading off performance
and resource overheads with vulnerability to
data races and recovery latency.
Critically, the response latency is not determined
by the frequency of checkpointing,
and sub-millisecond added delay is achieved with
all our server applications.
As we have found (\S\ref{subsec:eval_recovery}),
legacy applications may still have data races.
\name targets data races that
are most likely to remain undetected and uncorrected,
namely, rarely-manifested data races.
Unlike mechanism based strictly on active replication
and deterministic replay~\cite{rex}, \name is not affected by
data races that manifest during normal operation,
long before failure.
For handling data races that manifest right before failure,
\name introduces a simple best effort mechanism that
significantly reduces of the probability of the
data races causing recovery failure.
\name is a full fault tolerance mechanism.
It can recover from primary or backup host failure
and includes transparent failover of TCP connections.

This paper describes key implementation challenges
encountered in the development of \name and outlines
their resolution.
The extensive evaluation of \name{}, based on
eight benchmarks,
included performance and resource overheads,
impact on response latency, as well as recovery rate and latency.
The recovery rate evaluation, based on fault injection,
subjected \name to particularly harsh conditions
by intentionally perturbing the scheduling on the primary,
thus challenging the deterministic
replay mechanism (\S\ref{sec:exp}).
With high checkpointing frequency (\namese{}),
\name{}'s throughput overhead is less than 68\%
for seven of our benchmarks and 145\% for the eighth.
If the applications are known to be data race free,
with a lower checkpointing frequency (\namele{}),
the overhead is less than 59\% for all
benchmarks, significantly outperforming NiLiCon~\cite{nilicon}.
With data race free applications, \name recovered
from all fail-stop failures.
With two applications with
infrequently-manifested data races,
the recovery rate was over 99.4\% with \namese{}.
 \bibliographystyle{plain}
\bibliography{refs}
 
\end{document}